\documentclass[twocolumn,showpacs,preprintnumbers,amsmath,amssymb,prl]{revtex4}
\usepackage[dvips]{graphicx}
\begin{document}
\title{Unusual Metallic Conductivity of Underdoped and Optimally Doped
Cuprates: Evidence for Competing Fermi-Liquid and Pairing Pseudogap Effects}
\author{S. Dzhumanov and E. Dushanov}
\address{Institute of Nuclear Physics, 702132 Tashkent, Uzbekistan}
\begin{abstract}
        We propose a possible scenario for the new metallic conductivity of
      underdoped and optimally doped cuprates. Charge carriers are
      assumed to be large polarons which form a Fermi-liquid and Cooper-like
      pairs below a crossover tempurature $T^{\ast}$. We use the Boltzmann
      equation to calculate the conductivity of self-trapped carriers and the
      resistivity $\rho$ as a function of temperature and doping for different
      cuprates. We show that various anomalies in $\rho(T)$ below $T^{\ast}$
      are caused by the competing Fermi-liquid and BCS-like precursor pairing
      effects. Our results for $\rho$ fit well with existing experiments and
      characterize high-$T_c$ cuprates with an intermediate-coupling.

\end{abstract}
\pacs{71.38. + i; 72.10. -- d; 74.20. Fg; 74.72. -- h}
\maketitle

Understanding the normal-state charge transport properties of the high-$T_c$
cuprates  remains  one  of  the central issues  in  condensed  matter
physics \cite{1,2,3,4}. Many experimental studies \cite{5,6,7,8} have shown
that in the underdoped and optimally doped cuprates, the resistivity $\rho$
shows anomalous temperature dependences as well as a complicated doping
dependence. There is much evidence for the crossover regime at some
temperature $T^{\ast}>T_c$ in these materials and $\rho$ shows a
$T$-linear dependence above $T^{\ast}$.  However, for the underdoped cuprates
$\rho(T)$ starts to deviate downward from the $T$-linear behavior below
$T^{\ast}$. In contrast, in the optimally doped cuprates, $\rho(T)$ is roughly
linear below $T^{\ast}$. Sometimes the anomalous resistive transitions (i.e., a
sharp drop and small jump in $\rho(T)$) are observed at $T^{\ast}$ in some
cuprates (see, Refs. \cite{9,10}). The anomalies in $\rho(T)$ at $T\le
T^{\ast}$ are thought to arise from a pseudogap (PG) state which has been
observed by NMR, ARPES and other experiments \cite{11,12,13}. The origin of the
PG is still controversial. A number of theoretical models have been proposed,
which rely on different non-phononic mechanisms of pairing (including precursor
superconducting (SC) fluctuations) \cite{14,15}. On the other hand, there
is convincing experimental evidence for a strong electron-phonon interaction in
the cuprates \cite{13,16,17}. One possible scenario involving
electron-phonon interactions is based on the BCS-like non-SC (i.e.,
precursor) pairing model \cite{18}. It was postulated in this model that the
formation of the non-SC Cooper-like polaron pairs is quite possible in the
normal state, while their condensation into a superfluid Bose liquid state
would occur only at $T_c$. Further, it was argued that the pairing PG and true
SC gap have different origins and coexist below $T_c$. Recent experimental
results \cite{19,20} support such a picture and put much more severe
constraints on theories of precursor SC fluctuations. The opening of
the pairing PG in the normal state of the cuprates should affect their
transport properties. But the effect of the BCS-like non-SC gap (or PG) on the
charge transport in cuprates has not previously been studied.  Another open
question is the relevance of a Fermi-liquid picture to the normal state of the
underdoped and optimally doped cuprates \cite{4,13}.

So far, the $T$-linear resistivity in the cuprates have been explained in terms
of non-Fermi-liquid (including the RVB \cite{21}, marginal Fermi-liquid
\cite{22} and bipolaronic \cite{23}) models and different Fermi-liquid
scenarios (see Ref. \cite{3}). However, rather little is known theoretically
\cite{3,24} about the above anomalies in $\rho(T)$ and how the electron-phonon
coupling, Fermi-liquid and BCS-like precursor pairing correlations influence on
the normal-state charge transport properties of high-$T_c$ cuprates.

In this letter, we address these questions and propose a possible scenario for
the new metallic conductivity of underdoped and optimally doped cuprates. A key
is that the doped carriers in these materials above $T^{\ast}$ are the
non-interacting large polarons,while below $T^{\ast}$ they form a polaronic
Fermi-liquid and Cooper-like pairs. These large polarons are scattered by
acoustic phonons. We argue that the $T$-linear resistivity in underdoped and
optimally doped cuprates above $T^{\ast}$ is due to the carrier-phonon
scattering. We then show how the competing Fermi-liquid and BCS-like precursor
pairing correlations lead to the characteristic deviations from the $T$-linear
behavior in $\rho(T)$ below $T^{\ast}$ and the distinctly different
resistive transitions at $T^{\ast}$, which are very similar to the existing
experimental data on cuprates. We find that the Fermi-liquid parameter $F^s_1$
decreases as the BCS-like pairing PG or $T^{\ast}$ grows towards underdoped
regime.

{\it Relevant charge carriers in doped cuprates.} --- The undoped cuprates
are the charge-transfer (CT)-type Mott insulators. Upon doping the oxygen
valence band of the cuprates is occupied by holes. These charge carriers
being placed in a polar crystal will interact with the acoustic and optical
phonons, and the ground states of the doped carriers interacting with lattice
vibrations are their self-trapped (polaronic) states lying in the CT gap of
cuprates. Theoretical \cite{25} and experimental \cite{26} studies show that
the relevant charge carriers in doped cuprates are polaronic quasiparticles and
the electron-phonon interaction is responsible for the enhanced polaron mass
$m_p=2-2.5m_e$ \cite{26,27} (where $m_e$ is the free electron mass).

{\it Large-polaron transport above $T^{\ast}$.} --- As the doping $x$ (e.g., in
La$_{2-x}$Sr$_x$CuO$_4$) is increased towards the underdoped level, the
hole-doped cuprates evolve towards intermediate electron-phonon coupling regime
and enter the strange metallic state. We assume that the doped carriers in the
metallic state of high-$T_c$ cuprates are degenerate large polarons and form a
non-interacting Fermi-gas above $T^{\ast}$. We calculate the conductivity of
these polarons using the Boltzmann equation in the relation time
approximation. The Fermi energy of large polarons $\varepsilon_F=(\hbar^2/2m_p)
(3\pi^2n_p)^{2/3}$ (where $n_p$ is the concentration of large polarons) is of
order 0.1 eV or greater and the condition $\varepsilon_F\gg k_BT$ is satisfied
even well above $T^{\ast}$. We argue that the scattering of large polarons by
acoustic phonons determines their transport relaxation rate which depends on
the polaron energy $\varepsilon$ and is linear in temperature \cite{28}:
\begin{equation}
\frac{1}{\tau(\varepsilon)}=\frac{E^2_d(2m_p)^{3/2}\varepsilon^{1/2}k_BT}{
\pi\rho_M\hbar^4v^2_s},
\end{equation}
where $E_d$ is the deformation potential, $\rho_M$ is the density of the
materials, $v_s$ is the sound velocity.

The conductivity $\sigma$ is then calculated as
\begin{equation}
\sigma=-\frac{2n_pe^2}{3m_p}\cdot\frac{\int\varepsilon^{3/2}\tau(\varepsilon)
(\partial f_0/\partial\varepsilon)d\varepsilon}{\int\varepsilon^{1/2}f_0(
\varepsilon)d\varepsilon},
\end{equation}
where $f_0(\varepsilon)=[1+\exp(\varepsilon/k_BT)]^{-1}$ is the Fermi
distribution function. The integrals in Eq. (2) can be evaluated using the
standard approximations --- $\partial f_0/\partial\varepsilon=\delta(\varepsilon
-\varepsilon_F), f_0(\varepsilon<\varepsilon_F)=1$ and
$f_0(\varepsilon>\varepsilon_F)=0.$ Hence, for $T>T^{\ast}$ the conductivity is
given by
\begin{equation}
\sigma(T)=\frac{2\pi\hbar^4n_pe^2\rho_Mv_s^2}{E_d^2(2m_p)^{5/2}
\varepsilon^{1/2}_Fk_BT}.
\end{equation}

{\it Fermi-liquid and BCS-like precursor pairing correlations below
$T^{\ast}$.} --- We now turn to the Fermi-liquid and BCS-like precursor
pairing scenarios below $T^{\ast}$. We now make the key assumption that the
large polarons begin to form an interacting Fermi-liquid and non-SC Cooper-like
polaron pairs below $T^{\ast}$. In this case, in addition to the contribution
of the electron-phonon interaction to the carrier effective mass, there is a
contribution from the carrier-carrier interaction. The polaron mass $m_p$ is
then altered to $m^{\ast}_p=m_p(1+\frac13F^s_1$), where $F^s_1$ is the Landau
parameter. In principle, the BCS-like pairing of carriers may occur, depending
on the electron-phonon coupling strength, not only at $T_c$ but also far above
$T_c$ \cite{18}. In the weak-coupling regime, the pure dynamic-phonon-mediated
BCS pairing occurs at $T_c$. While in the strong-coupling limit, the
static-phonon-mediated non-BCS pairing occurs in the real space. Hence, the
combined static- and dynamic-phonon-mediated BCS-like pairing at some
temperature $T^{\ast}>T_c$ should occur in the intermediate-coupling regime.
\begin{figure}
\includegraphics[width=83mm]{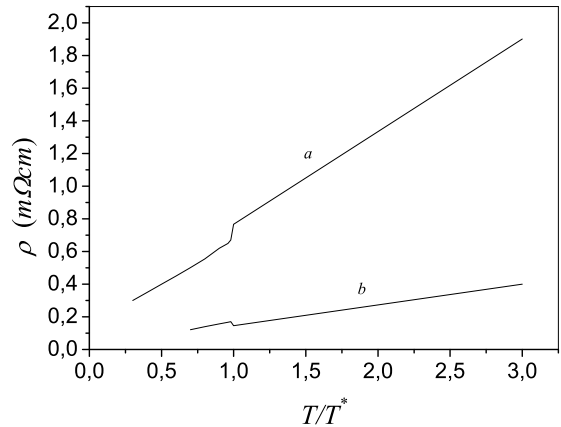}
\caption{Temperature dependence of $\rho$ calculated for an underdoped (curve
$a$) and optimally doped (curve $b$) system exhibiting distinctly different
Fermi-liquid and PG behaviors.} \end{figure}

A generalized BCS-like formalism, applied to self-trapped quasiparticles with a
well defined density of states $D(\varepsilon)$, leads to the following gap (or
PG) equation:
\begin{equation}
\Delta_F(k,T)=
-\sum\limits_{k'}V_{kk'}\frac{\Delta_F(k',T)}{2E(k',T)}\tanh\frac{E(k',T)}{2k_BT},
\end{equation}
where $E(k,T)=\sqrt{\varepsilon^2(k)+\Delta^2_F(k,T)}, V_{kk'}$ is the
pair interaction potential between polarons approximated as \cite{18}
\begin{equation}
V_{kk'}=\left\{\begin{array}{cl}
V_C-V_A,&0\le\varepsilon(k),\varepsilon(k')\le\varepsilon_A,            \\
V_C     &\varepsilon_A\le\varepsilon(k), \varepsilon(k')\le\varepsilon_C,\\
0       & {\rm otherwise,}
\end{array}\right.
\end{equation}
with the cut-off energies $\varepsilon_A=E_{bB}+\hbar\omega$ and
$\varepsilon_C\gg\varepsilon_A$ for the attractive $V_A$ and repulsive $V_C$
parts of $V_{kk'}$, respectively, $E_{bB}$ is the binding energy of a large
bipolaron (real space pairs), $\omega$ is a characteristic phonon frequency.
The solution of Eq. (4), which is obtained numerically using the model
potential (5), gives $\Delta_F(T)$. The PG temperature $T^{\ast}$ corresponding
to $\Delta_F(T^{\ast})=0$ is given by
$1.76k_BT^{\ast}=\Delta_F(0)=\varepsilon_A/\sinh(1/\lambda_{BCS})$, where
$\lambda_{BCS}=D(0)\tilde V$ is the BCS-like coupling constant,
$\tilde V=V_A-V_C/[1+D(0)V_C\ln(\varepsilon_C/\varepsilon_A)]$ is the effective
pairing potential. Below $T^{\ast}$, almost all polarons, which take part in
conduction, have the energies smaller than $\varepsilon_A$ and Fermi
distribution function has the form $f_0(\varepsilon)=[1+\exp(
\sqrt{\varepsilon^2+\Delta^2_F}/k_BT)]^{-1}$. The contribution of
polarons with $\varepsilon>\varepsilon_A$ to the conductivity is negligible
and can be ignored. Thus, the conductivity below $T^{\ast}$ is given by
%
\begin{equation}
\sigma(T)=\frac{A\cdot\!\int\limits^{\varepsilon_A}_0
\left\{f_0(\varepsilon)[1-f_0(\varepsilon)]/\sqrt{\varepsilon^2+
\Delta^2_F}\right\}\varepsilon^2d\varepsilon}{(k_BT)^2\int\limits^{
\varepsilon_A}_0f_0(\varepsilon)\varepsilon^{1/2}d\varepsilon},
\end{equation}
%
where
$A=4\pi\hbar^4n_pe^2\rho_Mv^2_s/3E^2_d\left[2m_p(1+\frac13F^s_1)\right]^{5/2}$.

{\it Competing Fermi-liquid and pairing pseudogap effects.} --- We use the
solution of Eq. (4) to calculate $\sigma(T<T^{\ast})$ by numerical integrating
Eq. (6). Here and below we use the fixed value $\varepsilon_A=0.08 eV$. Since,
$E_{bB}$ increases with increasing the electron-phonon interaction strength,
while $\omega$ decreases due to the phonon softening. Hence, $\varepsilon_A$
is nearly independent of doping. The Fermi energy for undoped cuprates is about
$E_F=7\ eV$ \cite{29} and $E_d$ is estimated as $E_d=(2/3)E_F$. The values
of $\rho_M$ and $v_s$ lie in the ranges $\rho_M\simeq 6-7\ g/cm^3$ and
$v_s\simeq(5-7)\cdot10^5 cm/s$ \cite{1,30}. Here we take $\rho_M=6\ g/cm^3$ and
$v_s=5\cdot10^5\ cm/s$.  To illustrate the competing Fermi-liquid and pairing
PG effects on the resistivity $\rho(T)=\rho_0+1/\sigma(T)$ (where $\rho_0$ is
the residual resistivity), we show in Fig. 1 results of our calculations for an
underdoped (curve {\it a}) and optimally doped (curve {\it b}) system obtained
using the parameters $m_p=2.2m_e, F^s_1=0.15, \lambda_{BCS}=0.67, T^{\ast}=250
K, n_p=0.46\cdot10^{21}cm^{-3}, \rho_0=0,2 m\Omega cm$ and $m_p=1.7m_e,
F^s_1=1.8, \lambda_{BCS}=0.49, T^{\ast}=140 K, n_p=0.85\cdot10^{21}cm^{-3},
\rho_0=0.02 m\Omega cm$, respectively. We find that the anomalies (i.e., a
sharp drop and small jump) in $\rho(T)$ reflect the competing Fermi-liquid and
PG effects below $T^{\ast}$. These anomalies in $\rho(T)$ are similar to the
resistive transitions observed above $T_c$ in some cuprates \cite{9,10}. The
Fermi-liquid effect is expected to disappear in the heavily underdoped regime.
\begin{figure}
\includegraphics[width=83mm]{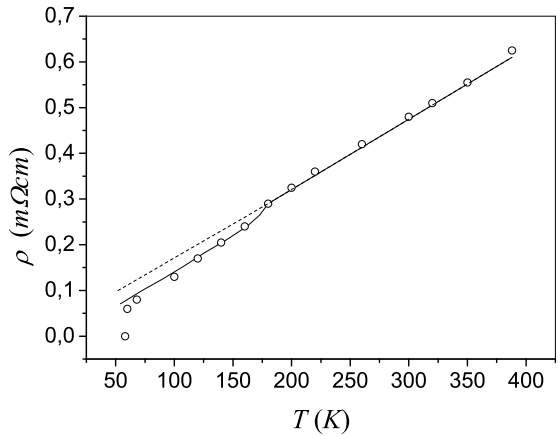}
\caption{The calculated temperature dependence of $\rho$ (solid line) compared
with experimental data on YBa$_2$Cu$_3$O$_{6.61}$ (taken from Ref. [5])
(open circles)}
\end{figure}
\begin{figure}
\includegraphics[width=83mm]{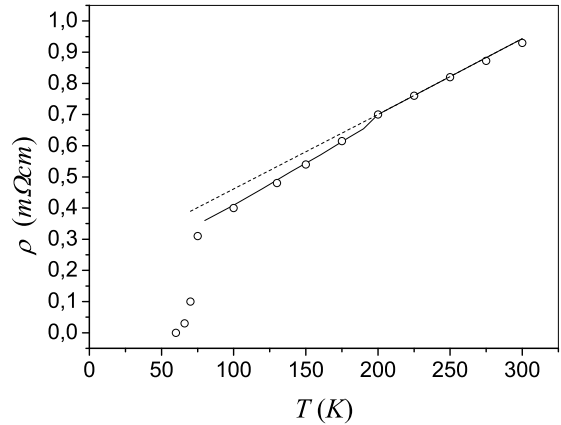}
\caption{The calculated temperature dependence of $\rho$ (solid line) compared
with experimental data on Bi$_2$Sr$_2$Ca$_{0.92}$Y$_{0.08}$Cu$_2$O$_8$ (taken
from Ref. [7]) (open circles)}
\end{figure}

{\it Comparison with existing experiments.} --- For the comparison with
existing experimental data we also present our results for $T$-dependent
resistivity in underdoped YBa$_2$Cu$_3$O$_{6.1}$,
Bi$_2$Sr$_2$Ca$_{0.92}$Y$_{0.08}$Cu$_2$O$_8$, La$_{1.92}$Sr$_{0.08}$CuO$_4$ and
optimally doped La$_{1.85}$Sr$_{0.15}$CuO$_4$ with the appropriate sets of
fitting parameters. Some relevant parameters were: (i) $m_p=2.3m_e, \rho_M=6.4\
g/cm^3$ and $v_s=6\cdot10^5cm/s$ for YBa$_2$Cu$_3$O$_{6.1}$; (ii) $m_p=2.2m_e,
\rho_M=6\ g/cm^3$ and $v_s=4.7\cdot10^5cm/s$ for Bi$_2$Sr$_2$Ca$_{0.92}$Y$_
{0.08}$CuO$_8$; (iii) $m_p=2.2m_e, \rho_M=6.1\ g/cm^3$ and $v_s=5\cdot10^5cm/s$
for La$_{1.92}$Sr$_{0.08}$CuO$_4$ and (iv) $m_p=1.8m_e, \rho_M=6.9\ g/cm^3$ and
$v_s=5.5\cdot10^5cm/s$ for La$_{1.85}$Sr$_{0.15}$CuO$_4$. Other fitting
parameters are presented in Table 1. The best fits are obtained (see Figs. 2-5)
if the combined Fermi-liquid and PG effects are taken into account. We believe
that the present theory describes well both the $T$-linear behavior of $\rho(T)$
above $T^{\ast}$ and the distinctly different deviations from the $T$-linear
behavior in $\rho(T)$ below $T^{\ast}$. One can see that the gradual
evolution of $\rho(T)$ with doping from a pronounced non-linear behavior in the
underdoped regime to the nearly $T$-linear behavior at some optimal doping is
associated with the competition between Fermi-liquid and PG effects. For larger
values of the PG $\Delta_F$, the Fermi-liquid parameter $F^s_1$ will be smaller.
%
%
\begin{table}
\squeezetable
\caption{Fitting parameters for underdoped and optimally doped cases. The
values of $T^{\ast}$ are taken from Refs. [5-8]}
\begin{tabular}{lcclcc}\hline\hline
Sample
       & $n_p,10^{21}$ & $F_1^s$ & $\lambda_{BCS}$ & $T^{\ast},$ & $\rho_0$       \\
       &$cm^{-3}$&&             & $K$         & $(m\Omega cm)$ \\
\hline
YBa$_2$Cu$_3$O$_{6.1}$
       & 0.52  & 0.60\hphantom{00}& 0.557   & 180        & 0.020          \\
Bi$_2$Sr$_2$Ca$_{0.92}$Y$_{0.08}$CuO$_8$
       & 0.50  & 0.57\hphantom{00}& 0.585   & 200        & 0.214          \\
La$_{1.92}$Sr$_{0.08}$CuO$_4$
       & 0.50  & 0.54\hphantom{00}& 0.585   & 200        & 0.210          \\
La$_{1.85}$Sr$_{0.15}$CuO$_4$
       & 0.80  & 1.44\hphantom{00}& 0.508   & 150        & 0.017          \\
\hline\hline
\end{tabular}
\end{table}
\begin{figure}
\includegraphics[width=83mm]{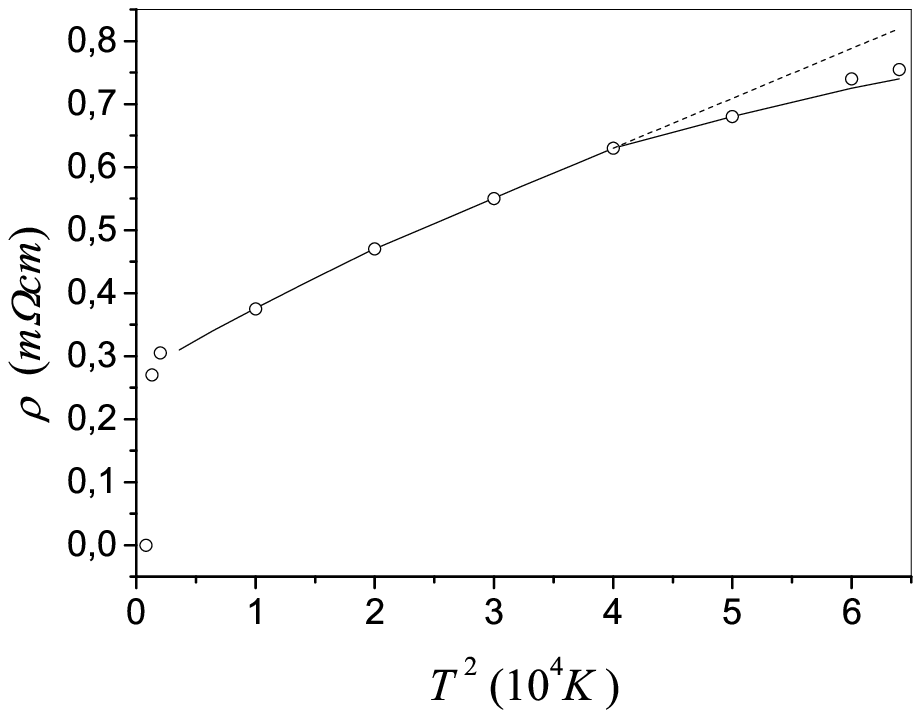}
\caption{The calculated temperature dependence of $\rho$ (solid line) compared
with experimental data on La$_{1.92}$Sr$_{0.08}$CuO$_4$ (taken from Ref. [8])
(open circles)}
\end{figure}
\begin{figure}
\includegraphics[width=83mm]{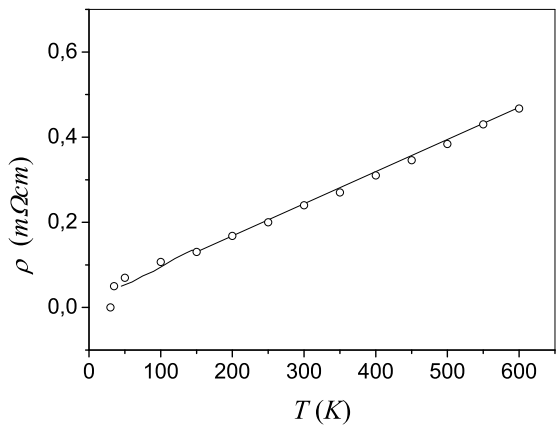}
\caption{The calculated temperature dependence of $\rho$ (solid line) compared
with experimental data on La$_{1.85}$Sr$_{0.15}$CuO$_4$ (taken from Ref. [6])
(open circles)}
\end{figure}

In conclusion, we have studied the unusual metallic conductivity of self-trapped
carriers (large polarons) in the normal state of underdoped and optimally doped
cuprates. These quasiparticles are assumed to form a noninteracting
polaronic gas above $T^{\ast}$ and an interacting polaronic Fermi-liquid below
$T^{\ast}$. We argue that the linear $T$-resistivity in underdoped and
optimally doped cuprates could actually be due to carrier-phonon scattering. We
have found that the distinctly different anomalies (e.g., a sharp drop and
small upturn, a gradual downward deviation from the $T$-linear behavior,
vanishing nonlinearity) in $\rho(T)$ occurring in these materials at $T^{\ast}$
or below $T^{\ast}$ are caused by the competing polaronic Fermi-liquid and
pairing PG effects. The proposed theory gives a good quantitative description
of the systematic evolution of $\rho(T)$ with doping observed in cuprates.

We thank A.S. Alexandrov, G. Baskaran, B. Batlogg, J.R. Cooper, A. Furrer,
E.M. Ibragimova, A. Junod, K.V. Mitsen and D. Singh for important discussions.
This work was supported by the Science and Technology Center of Uzbekistan and
in part by the Fundamental Research Foundation of Uzbek Academy of Sciences.

\end{document}